\documentclass[sigconf]{acmart}

\copyrightyear{2026}
\acmYear{2026}
\setcopyright{cc}
\setcctype{by}
\acmConference[CHI EA '26]{Extended Abstracts of the 2026 CHI Conference on Human Factors in Computing Systems}{April 13--17, 2026}{Barcelona, Spain}
\acmBooktitle{Extended Abstracts of the 2026 CHI Conference on Human Factors in Computing Systems (CHI EA '26), April 13--17, 2026, Barcelona, Spain}
\acmDOI{10.1145/3772363.3798580}
\acmISBN{979-8-4007-2281-3/2026/04}
\usepackage{booktabs}
\usepackage{tabularx,array}
\usepackage{placeins}

\begin{document}

\title{Input–Envelope–Output: \\Auditable Generative Music Rewards in Sensory-Sensitive Contexts}

\begin{abstract}
Generative feedback in sensory-sensitive contexts poses a core design challenge: large individual differences in sensory tolerance make it difficult to sustain engagement without compromising safety. This tension is exemplified in autism spectrum disorder (ASD), where auditory sensitivities are common yet highly heterogeneous. Existing interactive music systems typically encode safety implicitly within direct input–output (I–O) mappings, which can preserve novelty but make system behavior hard to predict or audit. We instead propose a constraint-first Input–Envelope–Output (I–E–O) framework that makes safety explicit and verifiable while preserving action–output causality. I–E–O introduces a low-risk envelope layer between user input and audio output to specify safe bounds, enforce them deterministically, and log interventions for audit. From this architecture, we derive four verifiable design principles and instantiate them in MusiBubbles, a web-based prototype. Contributions include the I–E–O architecture, MusiBubbles as an exemplar implementation, and a reproducibility package to support adoption in ASD and other sensory-sensitive domains.
\end{abstract}



\author{Cong Ye}
\authornote{Co-first author. Corresponding author.}
\affiliation{%
  \institution{College of Science, Mathematics and Technology, Wenzhou-Kean University}
  \city{Wenzhou}
  \state{Zhejiang}
  \country{China}
}
\email{1306248@wku.edu.cn}

\author{Songlin Shang}
\authornote{Co-first author}
\affiliation{%
  \institution{College of Science and Engineering, University of Minnesota}
  \city{Minneapolis}
  \state{Minnesota}
  \country{United States}
}
\email{shang217@umn.edu}

\author{Xiaoxu Ma}
\affiliation{%
  \institution{School of Electrical and Computer Engineering, Georgia Institute of Technology}
  \city{Atlanta}
  \state{Georgia}
  \country{United States}
}
\email{xma394@gatech.edu}

\author{Xiangbo Zhang}
\affiliation{%
  \institution{School of Mathematics, Georgia Institute of Technology}
  \city{Atlanta}
  \state{Georgia}
  \country{United States}
}
\email{xiangbo.zhang@gatech.edu}

\maketitle

\section{Introduction}

System builders creating interactive applications for sensory-sensitive populations face a fundamental design challenge: generative feedback can sustain engagement, but unconstrained generation risks sensory overload. This tension is particularly acute for individuals with autism spectrum disorder (ASD), where sensory processing differences are reported in over 90\% of the population~\cite{Tomchek2007Sensory,Leekam2007Sensory,BenSasson2009SensoryMeta,Chang2014Sensory}. Auditory sensitivities such as reduced sound tolerance, hyperacusis, and aversion to unpredictable sounds are particularly prevalent~\cite{Robertson2017Sensory}. Critically, these sensitivities are highly heterogeneous, making it difficult to design audio feedback that works for everyone.

Motor training is a cornerstone of daily intervention for children with ASD, supporting the development of coordination, motor planning, and functional skills~\cite{Sigrist2013Augmented,Perochon2023TabletGame}. Audio feedback plays a critical role in sustaining engagement during repetitive exercises~\cite{Sigrist2013Augmented,Danso2025PIMS}, and generative music offers promise by producing interaction-shaped outputs that maintain novelty while preserving causal relationships between actions and outcomes~\cite{PlutPasquier2020GenMusicGames,Vlist2011moBeat,Bergstrom2014MusicBiofeedback}. However, designing appropriate audio rewards presents a fundamental tension between constrained and unconstrained feedback. Constrained rewards, such as fixed or preset music, are predictable and safe but risk habituation as repeated playback diminishes users’ sense of agency~\cite{LopezDuarte2024PAMG,KojimaNakata2023RepetitiveListeningBoredom,Tabatabaie2014BoredomEEG}, whereas unconstrained generative rewards can sustain engagement~\cite{Danso2025PIMS,PlutPasquier2019MusicMatters} but introduce unpredictability that may trigger sensory overload in sensitive individuals~\cite{Robertson2017Sensory,Williams2021HyperacusisMeta}. Overall, most existing interactive music systems implement direct Input–Output (I–O) mappings, embedding safety-relevant constraints implicitly within musical rules rather than exposing them as auditable, configurable boundaries~\cite{Amershi2019Guidelines, PlutPasquier2020GenMusicGames}. This makes it difficult for system builders to tune the balance between habituation-prone constrained rewards and overload-prone unconstrained generation, leaving the effective Input–Output behavior opaque and potentially risky in sensory-sensitive contexts.

\begin{figure*}[!t]
  \centering
  \includegraphics[width=1.0\textwidth]{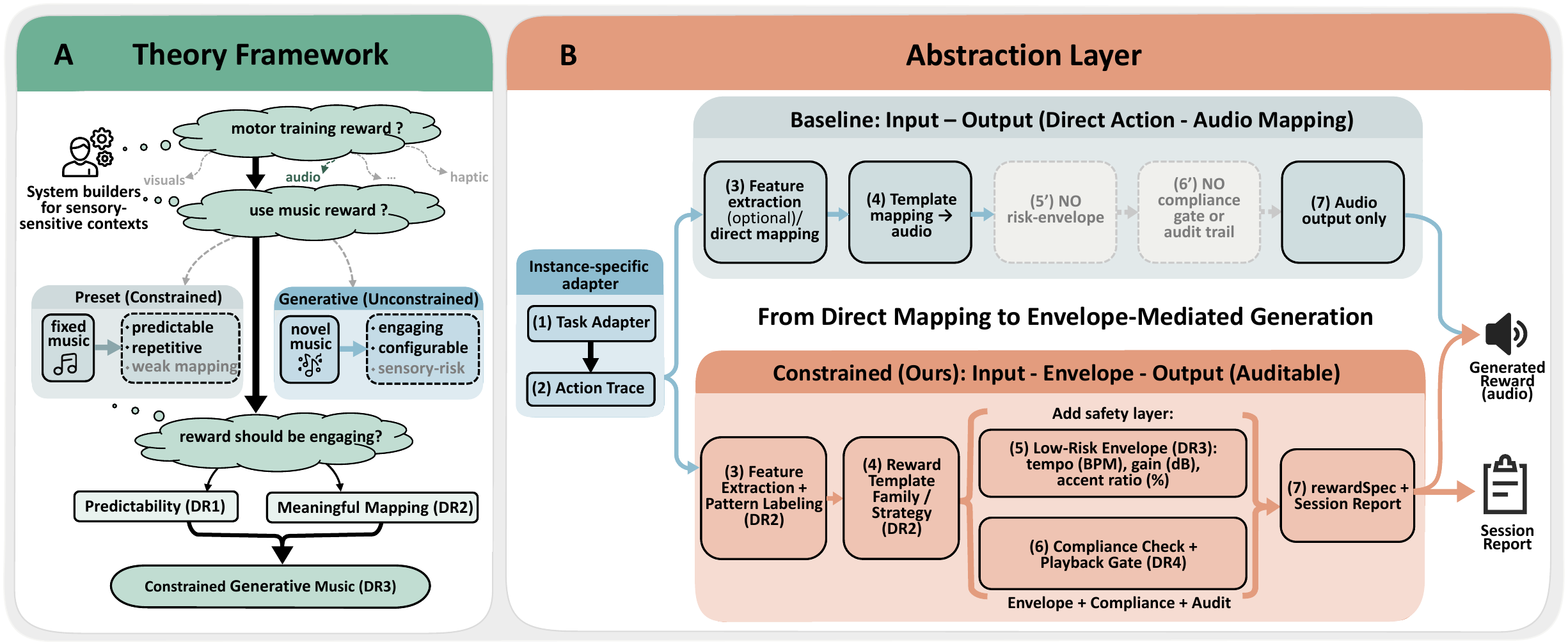}
  \Description{Diagram showing the design framework contrasting from Input–Output to Input-Envelope-Output architecture, with theory framework on the left and abstraction layer comparison on the right.}
  \caption{Design framework contrasting Input–Output to Input-Envelope-Output Approaches.
  (A) Theory Framework: design space leading to DR1–DR3.
  (B) Abstraction Layer: baseline direct mapping (top) vs our constraint-first pipeline (bottom) with explicit low-risk envelope, compliance gate, and auditable session report.}
  \label{fig:pipeline}
\end{figure*}

To address this gap, this paper presents a constraint-first generative reward framework that interposes an explicit low-risk envelope between user input and audio output (Fig.~\ref{fig:pipeline}). We use the term envelope in the systems sense—a declarative constraint layer that bounds audio generation parameters—rather than in the sound-synthesis sense of an ADSR amplitude curve. We term this the Input–Envelope–Output (I–E–O) paradigm, in contrast to conventional I–O mappings. The envelope declares conservative bounds on engine parameters (tempo, gain, accent ratio), enforces them via deterministic clamping, and logs all interventions for post-hoc audit. We formulate four design rationales grounding this approach as verifiable system requirements, and instantiate the framework in MusiBubbles, a web-based prototype for post-task music rewards in motor training.

We contribute: (1) four design rationales as transferable principles linking claims to system mechanisms and auditable evidence; (2) a reusable I–E–O framework architecture with explicit schemas and envelope enforcement; (3) MusiBubbles, a web-based reference implementation; and (4) a reproducibility package with synthetic traces, audit logs, and open-source artifacts.\footnote{Repository: \url{https://github.com/yecon-27/Input-Envelope-Output-Music}} Our primary target users are system builders (HCI/AI researchers, engineers, designers) creating generative reward feedback for sensory-sensitive contexts. Domain experts (music therapists, ASD practitioners, sensory-processing researchers) serve as gatekeepers who can inform envelope refinement in future ethics-approved studies. Children with ASD and caregivers are ultimate beneficiaries, but clinical effectiveness is out of scope.

\section{Related Work}

\paragraph{Music feedback in motor training} Audio feedback is widely used in motor learning to provide real-time performance information and sustain engagement during repetitive exercises~\cite{Sigrist2013Augmented}. Wang et al.~\cite{effectofmusic} used fixed-BPM music, set to the group’s pre-measured average tempo, as an audio reward/cue during multi-participant squat training. Perochon et al.~\cite{Perochon2023TabletGame} introduced a tablet-based bubble-popping game that extracts touch-derived motor features. Unlike prior work, where feedback is either so fixed that users’ engagement declines over time, or so variable that users might struggle to tolerate it and discern stable patterns, our method focuses on behavior-adaptive generative rewards.

\paragraph{Sensory-oriented design for ASD} Several interactive music systems target users with ASD~\cite{ChromeMusicLab, Ivanyi2024CADMI, karwankar2025ucue_idc}, and recent work has articulated sensory-oriented design principles for musical interfaces in ASD contexts~\cite{arora2025sensorynime}. Ivanyi et al.~\cite{Ivanyi2024CADMI} co-designed an accessible, collaborative digital musical interface with two special-education schools to support cooperative play and social-skill practice for children with ASD. Lee and Ho~\cite{multisensemusic} studied a multisensory music program for preschoolers with ASD and reported improved sensory processing. Prior systems hide safety constraints in musical rules, while our approach uses explicit, configurable bounds to accommodate highly diverse ASD preferences.

\paragraph{Constrained and auditable system design.} Interactive generative systems are moving toward auditability, yet Input–Output(I–O) constraints often remain implicit. Mitchell et al.~\cite{Mitchell2019ModelCards} and Gebru et al.~\cite{Gebru2021Datasheets} assume systems expose a clear, directly controllable I–O parameter interface, so constraints can be traced through documentation. By contrast, Zheng et al.~\cite{nime2024_86} implicitly embed constraints in rules, obscuring when they intervene along the I–O path. We instead implement I–E–O by inserting an explicit envelope (E) between Input and Output: it deterministically clamps parameters and logs both the requested value and the value actually used, making behavior predictable, verifiable, and auditable.

\section{Design Rationales}


In real-world sensory-sensitive settings, generative audio should prioritize controlled, inspectable variation over maximal expressiveness. We frame this as governing an uncertain Input--Output mapping, where small behavioral variability can trigger outsized perceptual swings. Based on this framing, we present four design rationales as verifiable system requirements, the bounds are conservative literature-informed priors, with human-subject validation left to future work. Table~\ref{tab:dr_evidence} links each rationale to its mechanism hook and engineering evidence.

\subsection{DR1: Predictability First.}

In sensory-sensitive contexts, the primary risk of generative audio is unexpected outputs with excessive variation~\cite{Robertson2017Sensory, Tomchek2007Sensory}. Even deterministic generators can amplify input variability into large swings in stimulation features, making overload risk difficult to anticipate~\cite{Wigham2015IU_Sensory,Boulter2014IU_ASD}. We define predictability as a deployable system property: output variation should be declaratively constrained, with identical inputs producing stable, reproducible results~\cite{Amershi2019Guidelines,NIST_AIRMF_100_1_2023}.

\subsection{DR2: Pattern-Level Mapping.}
Fixed rewards can lead to habituation, weakening the perceived action--outcome contingency~\cite{BeaudouinLafon2021Generative,Haggard2017SenseOfAgency}; in contrast, 1:1 sonification can translate behavioral noise into overly frequent acoustic events, increasing perceptual load~\cite{DubusBresin2013SonificationReview}. We target coarse pattern-level association: rewards reflect aggregate behavioral patterns rather than translating each action into an acoustic event, preserving interaction causality while leaving headroom for envelope constraints.

\subsection{DR3: Low-Risk Envelope.}
Unconstrained generation may produce high-stimulation features (loudness spikes, i.e., sudden increases in absolute output level measured in LUFS; dynamic jumps, i.e., abrupt transient changes in level between adjacent events measured as $\Delta$dB/s; and dissonant intervals) increasing sensory overload risk~\cite{Robertson2017Sensory, Takahashi2014Startle}. We operationalize risk mitigation via computable proxy metrics with declarative bounds~\cite{ITU_BS1770_5_2023,EBU_R128_2023}, enforced via deterministic clamping with all interventions logged~\cite{Schneider2000EnforceablePolicies,NIST_SP800_92_2006}.

\subsection{DR4: Auditable, Configurable, Reproducible.}
Sensory sensitivities are highly heterogeneous. Fixed thresholds can be too strict for some users and too permissive for others~\cite{Gajos2005Preference,Santos2024SMC}. We therefore design configurability as supervised, bounded, and auditable~\cite{Meyer1992DesignByContract,NIST_AIRMF_100_1_2023}: default mode uses conservative settings; tuning mode allows bounded exploration with automatic logging for replay and audit~\cite{Mitchell2019ModelCards,Raji2020AuditFramework}.

\begin{table}[!tbp]
\caption{Design rationales mapped to mechanism hooks and engineering evidence.}
\label{tab:dr_evidence}
\renewcommand{\arraystretch}{1.15}
\setlength{\tabcolsep}{5pt}
\small
\begin{tabularx}{\linewidth}{@{}
  >{\raggedright\arraybackslash}p{0.26\linewidth}
  >{\raggedright\arraybackslash}X
  >{\raggedright\arraybackslash}X @{}}
\toprule
\textbf{Design rationale (DR)} & \textbf{Mechanism hook} & \textbf{Engineering evidence} \\
\midrule
DR1: Predictability as deployable property & Explicit envelope; deterministic clamp/repair & Full bound compliance across all traces \\
DR2: Pattern-level mapping & Pattern labeling $\rightarrow$ template family & Onset density distributions distinguishable across pattern types \\
DR3: Low-risk envelope & Bound overload-prone dimensions; log interventions & Baseline vs constrained distribution shift; clamp rate summary \\
DR4: Auditable, configurable & Requested vs effective params; config hash + seed & Tuning monotonicity across Relaxed/Default/Tight configurations \\
\bottomrule
\end{tabularx}
\end{table}

\section{The I–E–O Framework}

We operationalize the four design rationales into the I–E–O framework, a pipeline with two key structures: an instance-specific layer and a transferable core. The \textbf{instance-specific layer} converts task-specific events into a standardized actionTrace schema with timestamped entries and notes; adapting to other discrete motor tasks only requires implementing a new adapter. The \textbf{transferable core} then processes actionTrace through four stages: (1) extract behavioral features and assign pattern labels (DR2); (2) select a reward-template family conditioned on the pattern label (DR2); (3) declare conservative bounds on engine parameters (DR3); and (4) enforce these bounds via deterministic clamping with audit logging (DR3/DR4). All generation parameters and enforcement decisions are logged in structured session reports, enabling full reproducibility. We instantiate the pipeline in MusiBubbles and evaluate it using non-human evidence artifacts.

\section{Prototype and Engineering Evaluation}

This section instantiates the Input-Envelope-Output(I–E–O) framework as a runnable prototype and evaluates it as an engineering system. Using MusiBubbles, we generate standardized action traces and compare an unconstrained baseline to the constrained pipeline under identical inputs and fixed seeds. We then report DR1–DR4 evidence artifacts—boundedness, clamp statistics, stimulation-proxy distribution shifts, and replayable session reports—to verify inspectability and reproducibility.

\subsection{MusiBubbles: Reference Implementation}

\begin{figure*}[!htbp]
    \centering
    \includegraphics[width=0.95\textwidth]{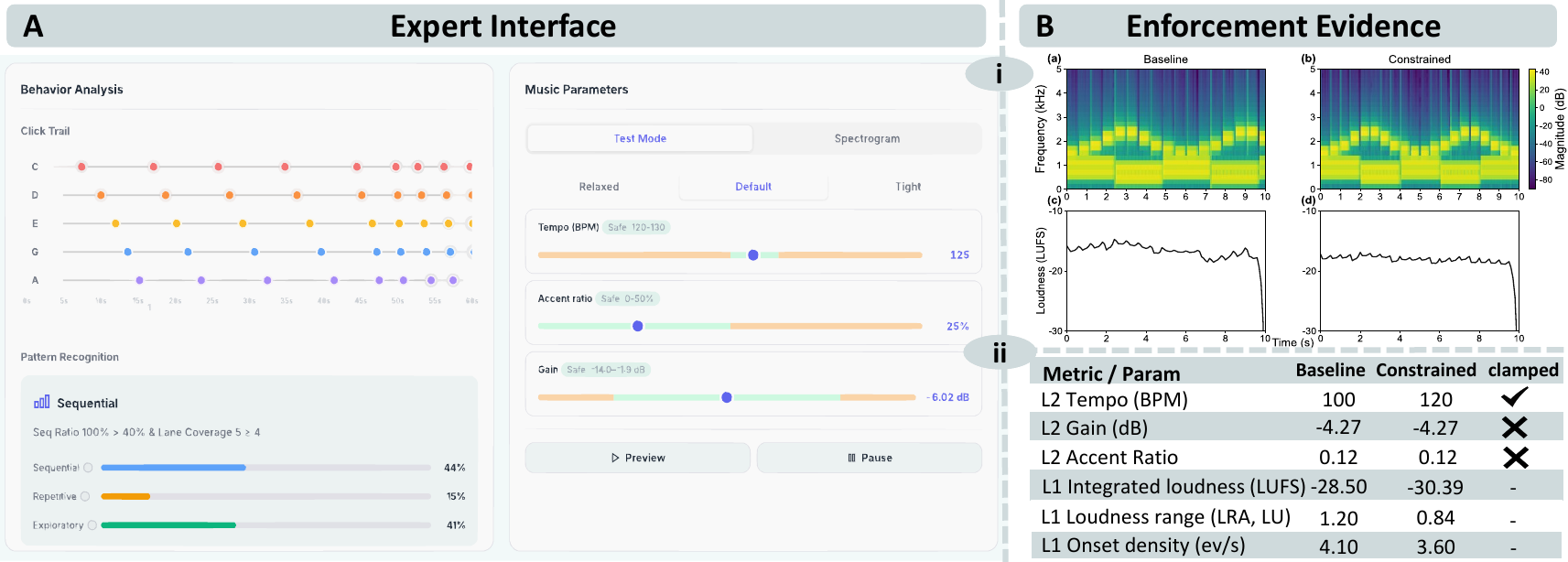}
    \Description{MusiBubbles interface showing expert mode with behavior analysis and music parameters configuration on the left; spectrogram comparison, LUFS loudness contours, and session report summary table on the right.}
    \caption{MusiBubbles reference implementation. (A) Expert interface: Behavior Analysis panel shows click trail and pattern recognition results; Music Parameters panel provides configurable Low-risk Range bounds (orange regions indicate out-of-bound zones) under Default configuration. (B) Enforcement evidence (under default) : (i) spectrogram comparison between baseline (a) and constrained (b) outputs showing differences in onset density, with corresponding loudness contours (LUFS) and range summaries (LRA) (c, d); spectrograms share the same color scale; (ii) summary table showing L2 parameters with baseline vs constrained values and clamp status, plus resulting L1 signal metrics. This single-trace example serves as diagnostic evidence; aggregate statistics (N=660) are reported in Figure~\ref{fig:tuning}.}
    \label{fig:musibubbles}
\end{figure*}

MusiBubbles is a reference implementation of I–E–O framework as shown in Fig.~\ref{fig:musibubbles}, designed to produce standardized actionTrace data, run baseline vs constrained comparisons, and generate structured session reports for engineering verification and reproducibility. It does not claim therapeutic efficacy or clinical safety.

The game presents a web-based bubble popping interface inspired by tablet motor training paradigms~\cite{Axford2018iPad,Perochon2023TabletGame}, featuring five vertical lanes mapping to C Major Pentatonic (C--D--E--G--A) to reduce dissonance risk~\cite{Frazee1987Orff}. The system includes an expert panel for real-time monitoring of session parameters and enforcement status. Each 60-second session records tap events into \texttt{actionTrace} entries containing timestamp, lane, intensity/outcome, and optional note fields.

Three pattern labels (Sequential, Repetitive, Exploratory) route traces to template families based on interpretable features: lane diversity, dominant lane ratio, and sequential coverage. When scores fall within 0.05, sequential priority applies as a deterministic tie-breaker. These are pragmatic routing labels, not validated behavioral constructs.

The low-risk envelope declares bounds on three engine parameters: tempo, gain (dB), and accent ratio (Table~\ref{tab:envelope}). These three dimensions are selected as the primary enforcement targets because they are directly computable before playback and map to auditable numeric values. Tempo and gain correspond to the stimulation dimensions most associated with auditory overload risk in ASD contexts---rhythmic rate and overall intensity~\cite{Robertson2017Sensory, Takahashi2014Startle}---while accent ratio serves as an engineering proxy for abrupt dynamic accentuation between events. The system applies deterministic clamping to all requested values, with each intervention logged in structured session reports to enable full replay and audit.

\subsection{Engineering Evaluation}

\textbf{Setup.} We evaluate envelope enforcement using N=660 synthetic \texttt{actionTrace} samples (see supplementary materials) under paired comparison design: each trace runs under baseline (no envelope) and constrained conditions with identical seeds. We test three configurations (Relaxed, Default, and Tight) as shown in Table~\ref{tab:configs} to demonstrate tuning sensitivity.

\begin{figure*}[!htbp]
    \centering
    \includegraphics[width=0.95\textwidth]{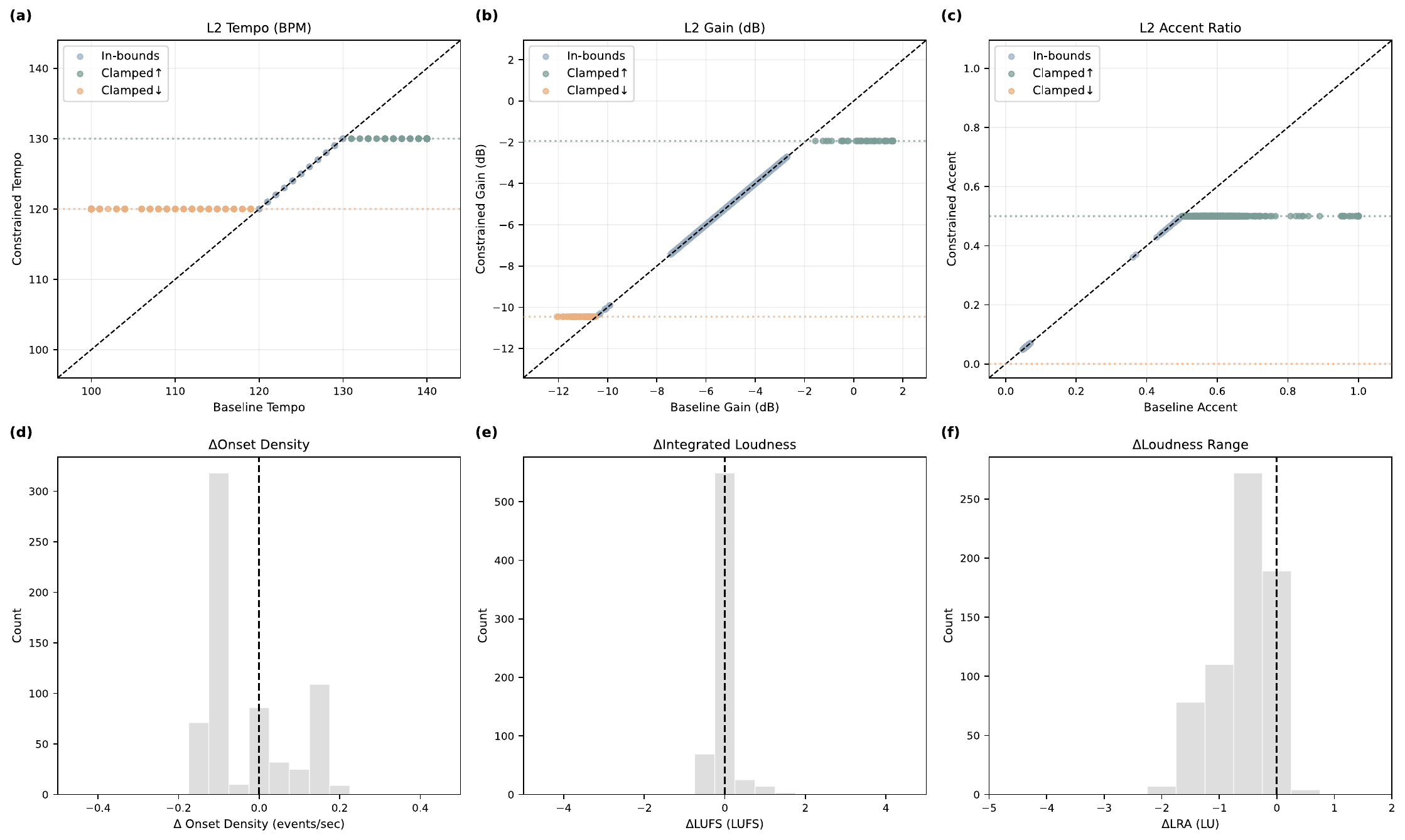}
    \Description{Six-panel figure showing envelope enforcement results: top row contains scatter plots for tempo, gain, and accent ratio comparing baseline vs constrained values; bottom row shows delta distributions for onset density, LUFS, and LRA.}
    \caption{Envelope enforcement under Default configuration (N=660 traces). Top row (a–c): scatter plots comparing baseline values (x-axis) with post-enforcement values (y-axis) for tempo, gain, and accent ratio; dashed line indicates y=x, dotted lines denote lower/upper bounds; clamp rates are 91.5\%, 18.0\%, and 85.6\%, respectively (gain shows lower clamp rate as user behavior rarely produces extreme values; accent ratio lower-bound clamping is rare as users seldom generate very low accent ratios). Bottom row (d–f): resulting distribution shifts in onset density, integrated loudness (LUFS), and loudness range (LRA); vertical dashed line indicates zero change. Additional configurations (Tight, Relaxed) are provided in supplementary materials.}
    \label{fig:tuning}
\end{figure*}

\textbf{L2 parameter verification (DR1, DR3).} The upper row of Fig.~\ref{fig:tuning} shows hockey-stick scatter plots for each L2 parameter. Points on the y=x diagonal (blue) were already within bounds; points deflected from the diagonal (orange) were clamped. Default configuration shows per-parameter clamp rates of 91.5\% (tempo), 18.0\% (gain), and 85.6\% (accent ratio), with 98.9\% of traces having at least one parameter clamped. Relaxed achieves only 8.9\% total clamp rate (59/660, all from Gain), confirming the enforcer does not over-intervene when bounds are wide. In all configurations, every post-enforcement value remains within its declared bounds.

\textbf{L1 signal diagnostics (DR4).} The lower row of Fig.~\ref{fig:tuning} shows $\Delta$ distributions (constrained $-$ baseline) for three L1 metrics: onset density, integrated loudness (LUFS), and loudness range (LRA). Distributions shift monotonically with constraint tightness: Tight produces the largest deviations from baseline, Default produces moderate shifts, and Relaxed distributions concentrate near zero. This demonstrates tuning monotonicity: tighter L2 bounds produce proportionally larger L1 effects, enabling practitioners to calibrate constraint strength.

\textbf{Pattern discriminability (DR2).} Onset density distributions remain distinguishable across pattern types even under Tight constraints, confirming that envelope enforcement does not flatten expressivity.

\begin{table}[!ht]
\caption{Two-layer envelope contract: Hard L2 parameters are enforced via clamping and audit-logged; Monitored L1 metrics are diagnostic evidence only.}
\label{tab:envelope}
\small
\renewcommand{\arraystretch}{1.1}
\begin{tabularx}{\columnwidth}{@{}p{0.22\columnwidth}p{0.14\columnwidth}X@{}}
\toprule
\textbf{Parameter} & \textbf{Layer} & \textbf{Evidence} \\
\midrule
\multicolumn{3}{@{}l}{\textit{Hard L2: Engine parameters (enforced + audit logged)}} \\
\midrule
Tempo & L2 (BPM) & Clamp rate; requested$\rightarrow$effective \\
Gain & L2 (dB) & Clamp rate; requested$\rightarrow$effective \\
Accent ratio & L2 (ratio) & Clamp rate; per-dimension distribution \\
\midrule
\multicolumn{3}{@{}l}{\textit{Monitored L1: Signal diagnostics (measured-only, not enforced)}} \\
\midrule
Integrated loudness & L1 (LUFS) & $\Delta$ distribution (constrained $-$ baseline) \\
Loudness range & L1 (LU) & $\Delta$ distribution; accent ratio effect \\
Onset density & L1 (ev/s) & $\Delta$ distribution; tempo effect \\
\bottomrule
\end{tabularx}
\end{table}

\begin{table}[!ht]
\caption{L2 parameter bounds under three configurations.}
\label{tab:configs}
\small
\renewcommand{\arraystretch}{1.1}
\begin{tabularx}{\columnwidth}{@{}lXXX@{}}
\toprule
\textbf{Parameter} & \textbf{Relaxed} & \textbf{Default} & \textbf{Tight} \\
\midrule
Tempo (BPM) & (60, 180) & (120, 130) & (124–126) \\
Gain (dB) & ($-60$, 0) & ($-10.5$, $-1.9$) & ($-6.9$, $-5.2$) \\
Accent ratio & (0.0, 1.0) & (0.0, 0.5) & (0.0, 0.1) \\
\bottomrule
\end{tabularx}
\end{table}

\section{Conclusion and Future Work}

We introduce Input–Envelope–Output, a constraint-first framework that makes generative feedback enforceable, auditable, and reproducible via declarative bounds, deterministic clamping, and replayable session reports. We instantiate it in MusiBubbles and validate engineering correctness with evidence artifacts, providing deployable infrastructure for controlled generative rewards and a foundation for future expert and human evaluations.

This study has two main limitations, both of which are deliberate design choices that sharpen the contribution. First, we validate engineering correctness rather than user outcomes: by proving that bounds are enforced, logged, and reproducible, we provide deployable infrastructure that can be reviewed and safely iterated before involving sensory-sensitive participants. Second, MusiBubbles uses only music feedback. New modalities will require modality-specific envelope dimensions, but the core contribution is still directly transferable. The enforcement and logging substrate remains modality-agnostic and carries over with declarative bounds and deterministic clamping. Future work will expand envelope dimensions, probe failure modes, and test broader tasks and musical traditions.

\begin{acks}
We thank Dr. Lingling Deng (Wenzhou-Kean University) for constructive feedback on an earlier draft, particularly comments that helped improve argument clarity and figure presentation. We also thank Bingyu Zhu (Registered Music Therapist, 
Melbourne, Australia) for consultation on music-therapy considerations, which informed our selection of musical parameters and feedback on early interface prototypes. The final manuscript was shaped by their generous input, though the authors alone are responsible for its final form.
\end{acks}

\clearpage
\bibliographystyle{ACM-Reference-Format}
\bibliography{references}

\end{document}